\documentclass[pra, aps, longbibliography, twocolumn, amsmath, amssymb, superscriptaddress,showpacs]{revtex4-1}
\usepackage{graphicx, array, booktabs, color, bm, upgreek}
\usepackage[colorlinks,linkcolor=blue,anchorcolor=blue,citecolor=blue,urlcolor=blue]{hyperref}

\begin{document}

\title{Revealing the working mechanism of quantum neural networks by mutual information}

\author{Xin Zhang}
\affiliation{College of Intelligence and Computing (School of Computer Science and Technology), Tianjin University, No.135, Ya Guan Road, Tianjin 300350, China}
\author{Yuexian Hou}
 \email{yxhou@tju.edu.cn}
\affiliation{College of Intelligence and Computing (School of Computer Science and Technology), Tianjin University, No.135, Ya Guan Road, Tianjin 300350, China}

\date{\today}

\begin{abstract}
Quantum neural networks (QNNs) is a parameterized quantum circuit model, which can be trained by gradient-based optimizer, can be used for supervised learning, regression tasks, combinatorial optimization, etc. Although many works have demonstrated that QNNs have better learnability, generalizability, etc. compared to classical neural networks. However, as with classical neural networks, we still can't explain their working mechanism well. In this paper, we reveal the training mechanism of QNNs by mutual information. Unlike traditional mutual information in neural networks, due to quantum computing remains information conserved, the mutual information is trivial of the input and output of $U$ operator. In our work, in order to observe the change of mutual information during training, we divide the quantum circuit ($U$ operator) into two subsystems, discard subsystem ($D$) and measurement subsystem ($M$) respectively. We calculate two mutual information, $I(Di:Mo)$ and $I(Mi:Mo)$ ($i$ and $o$ means input or output of the corresponding subsystem), and observe their behavior during training. As the epochs increases, $I(Di:Mo)$ gradually increases, this may means some information of discard subsystem is continuously pushed into the measurement subsystem, the information should be label-related. What's more, $I(Mi:Mo)$ exist two-phase behavior in training process, this consistent with the information bottleneck anticipation. The first phase, $I(Mi:Mo)$ is increasing, this means the measurement subsystem perform feature fitting. The second phase, $I(Mi:Mo)$ is decreasing, this may means the system is generalizing, the measurement subsystem discard label-irrelevant information into the discard subsystem as many as possible. Our work discussed the working mechanism of QNNs by mutual information, further, it can be used to analyze the accuracy and generalization of QNNs.

\end{abstract}

\maketitle

\section{introduction}
Many works have revealed that QNNs have advantages over classical neural networks in terms of performance \cite{1, 2, 3, 4, 5}, generalization \cite{6, 7, 8} and trainability \cite{9, 10}, but it also faced the problem of poor interpretability, we can't understand its decision-making process well. This limit its application in critical areas such as medical diagnostics and smart finance.

The information bottleneck based on mutual information has achieved some success in the interpretability of classical neural networks, the information bottleneck itself is based on rigorous mathematical theory \cite{11}. In recent years, Tishby $et\:al.$ have found that it can be used to explain deep neural networks, and discover information processing inequalities (Markov chain) for deep neural networks, the representation $T$ maximizes the compressed input $X$, and maximally preserves the mutual information about label $Y$ \cite{12}. And they found that there were two phases in the training process: fitting and compression, in fitting phase, the mutual information of the representation $T$ and input $X$ increases, the neural network fits the input data, in compression phase, the mutual information of the representation $T$ and input $X$ decreases, the neural network discard information that is not relevant to the label \cite{13}. Tishby’s work sheds some light on how neural networks work from an information perspective, but can it be used to explain QNNs? This is an important and open question.

Banchi $et\:al.$ first used information bottlenecks for quantum neural networks \cite{14}, they define three spaces: quantum state space $Q$, classical parameter space $X$ and class space $C$, and their mutual information $I(C:Q)$ and $I(X:Q)$, and the Lagrangian of the quantum information bottleneck is defined by mutual information $L_{IB}=I(X:Q)-\beta I(C:Q)$, the trade-off between accuracy and generalization is achieved by minimizing the Lagrangian for a certain value of $\beta$. A small $\beta$ means that the $X$ have maximum compression, the model generalizes well, a large $\beta$ means that $Q$ reserve more information related to $C$,the model have better accuracy.

The significance of paper \cite{14} is the first introduction of information bottlenecks into QNNs, and used it as a tool to analyze the accuracy and generalization of the QNNs, but their work is still not consummate, firstly, they didn't connect the information bottleneck to the interpretability of the QNNs Ansatz, secondly, they did not study the behavior of mutual information during training and did not find the critical two-phase phenomenon.

In paper \cite{15}, Zhai $et\:al.$ studied the magnetic properties of spin models using QNNs combined with tripartite mutual information, find a two-phase behavior during training, that is the tripartite mutual information increase first then decrease. They believe that the phase of tripartite mutual information increasing is learning small-scale structure, and the phase of tripartite mutual information decreasing is learning large-scale structure.

This paper focuses on the mutual information of QNNs Ansatz in the training process, thereby elaborates the working mechanism. Since the information processing of QNNs is unitary and the information is conserved, thus we cannot directly calculate the mutual information of input and output. In this paper, we borrow from scrambling model \cite{16}, we divide Ansatz into measurement subsystem and discarding subsystem, and calculate the mutual information behavior intra- and inter- subsystems separately, we find a two-phase behavior of the measurement subsystem in which mutual information increase first then decrease, the two-phase behavior may have different mechanisms with paper \cite{15}, because our measurement subsystem only have one qubit, it cannot learn large-scale structure, so we think it consistent with information bottlenecks anticipation. Moreover, in the framework of the information bottleneck, using the scheme of paper \cite{14}, we can study the accuracy and generalization error bounds of QNNs.

\section{Quantum neural networks model}
QNNs is a class of quantum circuits that can be used for classification and regression based on variational methods, which can be divided into three categories: explicit quantum models \cite{10, 17, 18}, quantum kernel method \cite{19, 20, 21} and data reuploading models \cite{22, 23}. Jerbi $et\:al.$ reviewed these three methods and analyzed their relationship in paper \cite{24}.

\begin{figure}
\includegraphics[width=7.5cm]{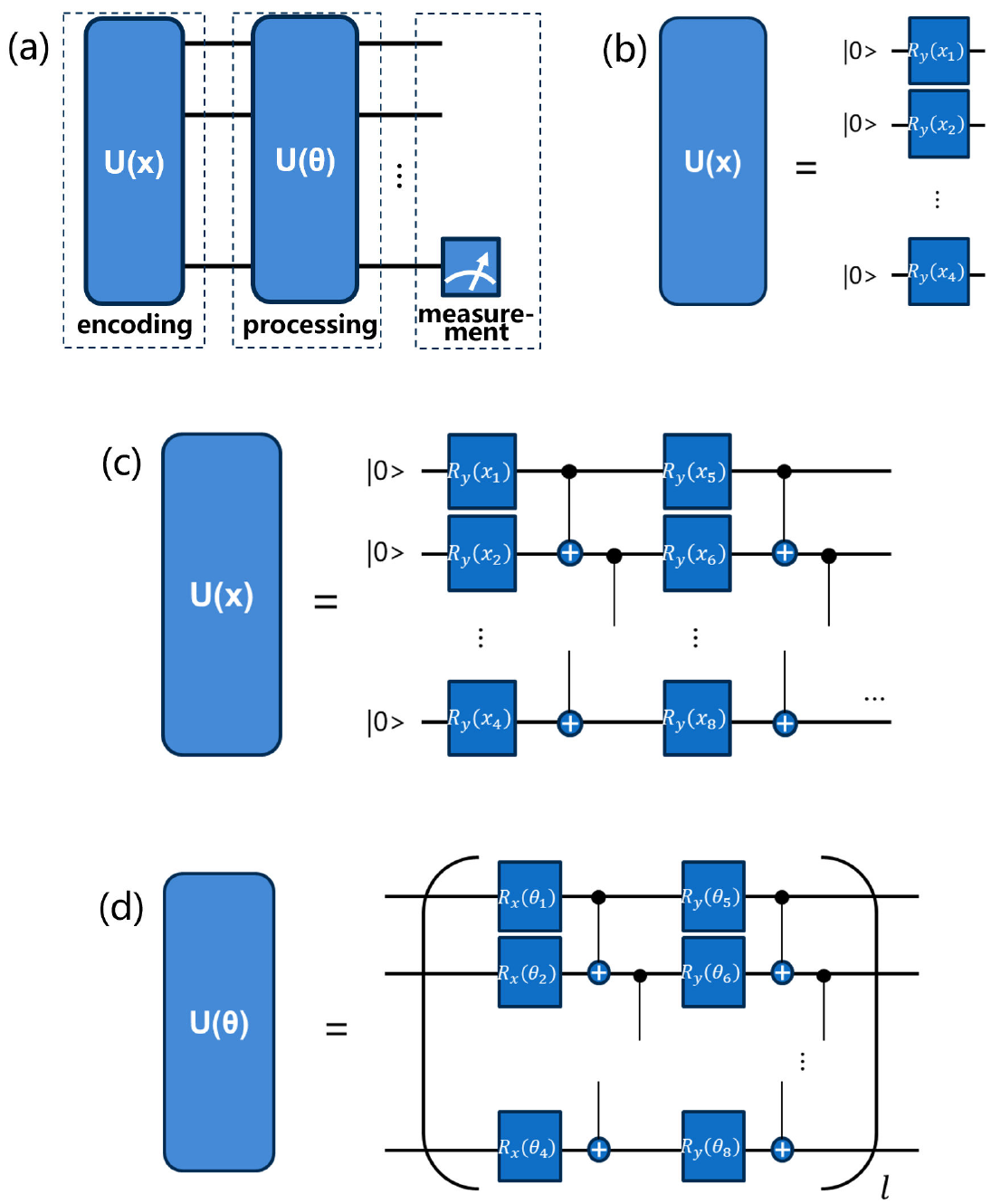}
\caption{Structure diagram of QNNs. (a)The overall architecture of QNNs, it contains three parts encoding, processing and measurement. (b)Illustrate qubit encoding method. (c)Illustrate Ansatz structure of Brick Wall, $l$ is the number of repetitions of the module. }
\label{fig:fig1}
\end{figure}

In this paper, we focus on explicit quantum models, it is also the most studied model at present, the overall architecture as shown in Fig. 1(a). This model can be divided into three parts, the first part is the data encoding, which mapped classical data (feature data) to quantum states, the commonly used methods are amplitude encoding and qubit encoding \cite{25}. Amplitude encoding is the mapping of feature data to probability amplitudes of quantum states, the encoding complexity is exponential. For qubit encoding method, each feature corresponds to a rotation gate, which shown in Fig. 1(b), and we can also use "interleaved" qubit encoding, which shown in Fig. 1(c), we can encode more features in fewer qubits, the encoding complexity is linear. The second part is state processing, the task is to evolve the encoded quantum state to a target quantum state through unitary transformation, that can minimize the value of the loss function, this is achieved by training the parameterized quantum circuits (Ansatz). We can build different Ansatz, this paper adopts Brick Wall structure, which shown in Fig. 1(d). Ansatz is the most important part of QNNs, and the interpretability of this paper is focused on it. The third part is measurement, we can get classification or regression result after measurement.

Suppose $\bold x_{\bold i}$ is the $i$th training sample, it is a $d$-dimensional vector ($d$ is the number of feature), $\bold y$ is the label corresponding to the sample (represented by a vector of 0,1), They form the sample pair ($\bold x_{\bold i}, \bold y$). $\bold x_{\bold i}$ is encoded as quantum state $\rho(\bold x_{\bold i})$ by qubit encoding, and the quantum state evolved into $U(\bm \uptheta)\rho(\bold x_{\bold i})U^\dagger(\bm \uptheta)$ after Ansatz, the result after measurement is $h_{k}(\bold x_{\bold i}, \bm{\uptheta})=Tr(M^\dagger_{k}M_{k}U(\bm \uptheta)\rho(\bold x_{\bold i})U^\dagger(\bm \uptheta))$ ($M_{k}$ is the Pauli $Z$ measurement, which  act on $k$th qubit), we can define the cross-entropy (CE) loss function,

\begin{eqnarray}\label{ME1}
&&L_{CE}(h(\bold x_{\bold i},\bm\uptheta),\bold y)=\sum_{k}y_{k}log(h_{k}(\bold x_{\bold i}, \bm{\uptheta}))
\end{eqnarray}

By taking the derivative of the loss function with $\bm{\uptheta}$, we obtain the gradient, but it's very different from classical neural network, we don't use back propagation to calculate the gradient, the generally used method is parameter shift \cite{26, 27}, the formula is as follows.

\begin{eqnarray}\label{ME1}
&&\frac{\partial L_{CE}(h(\bold x_{\bold i},\bm\uptheta),\bold y)}{\partial \theta}=\sum_{k}\frac{y_{k}}{h_{k}(\bold x_{\bold i}, \bm{\uptheta})}\frac{\partial h_{k}(\bold x_{\bold i}, \bm{\uptheta})}{\partial \theta}
\end{eqnarray}

\begin{eqnarray}\label{ME1}
&&\frac{\partial h_{k}(\bold x_{\bold i}, \bm{\uptheta})}{\partial \theta} \approx \frac{h_{k}(\bold x_{\bold i}, \theta + \frac{\Delta \theta}{2}) - h_{k}(\bold x_{\bold i}, \theta - \frac{\Delta \theta}{2})}{\Delta \theta}
\end{eqnarray}

After obtaining the gradient, we can update the trainable parameters $\bm{\uptheta}$ by optimizer such as gradient descent or Adam \cite{28}, until the loss function converges.

\section{Mutual information in QNNs}
In 2000, Tishby first proposed the information bottleneck theory based on mutual information, gave the mathematical definition and iterative method, and proved the convergence \cite{11}. With the rise of deep learning, Tishby $et\:al.$ used it to explain deep learning, they discovered the information processing inequalities in deep learning, and two-phases phenomenon in the training: fitting then compression \cite{12, 13}.

\begin{figure}
\includegraphics[width=5.5cm]{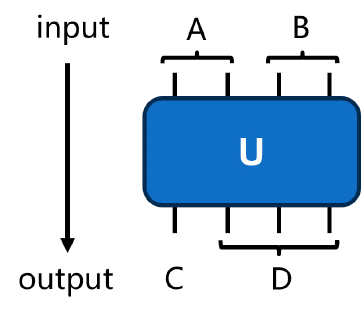}
\caption{Schematic diagram of the Scrambling model. First, we divided the input and output of $U$ operator into subsystems, such $A$, $B$, $C$ and $D$ in this figure. Next, we can calculate the mutual information of different subsystem, such as $I(A:C)$ or $I(B:C)$, thus we can obtain the property of $U$ operator by this mutual information. }
\label{fig:fig2}
\end{figure}

However, there is no sense calculating the mutual information of the input and output for QNNs, because the quantum information processing is unitary, there is no loss of information, the mutual information is constant. Scrambling model \cite{15} proposed by Hosur $et\;al.$ give the solution idea, we can divide the input and output of unitary evolution into different subsystems, we can obtain the property of $U$ by calculating the mutual information of different subsystems. The schematic diagram of the Scrambling model is shown in Fig. 2.

\begin{figure}
\includegraphics[width=7.5cm]{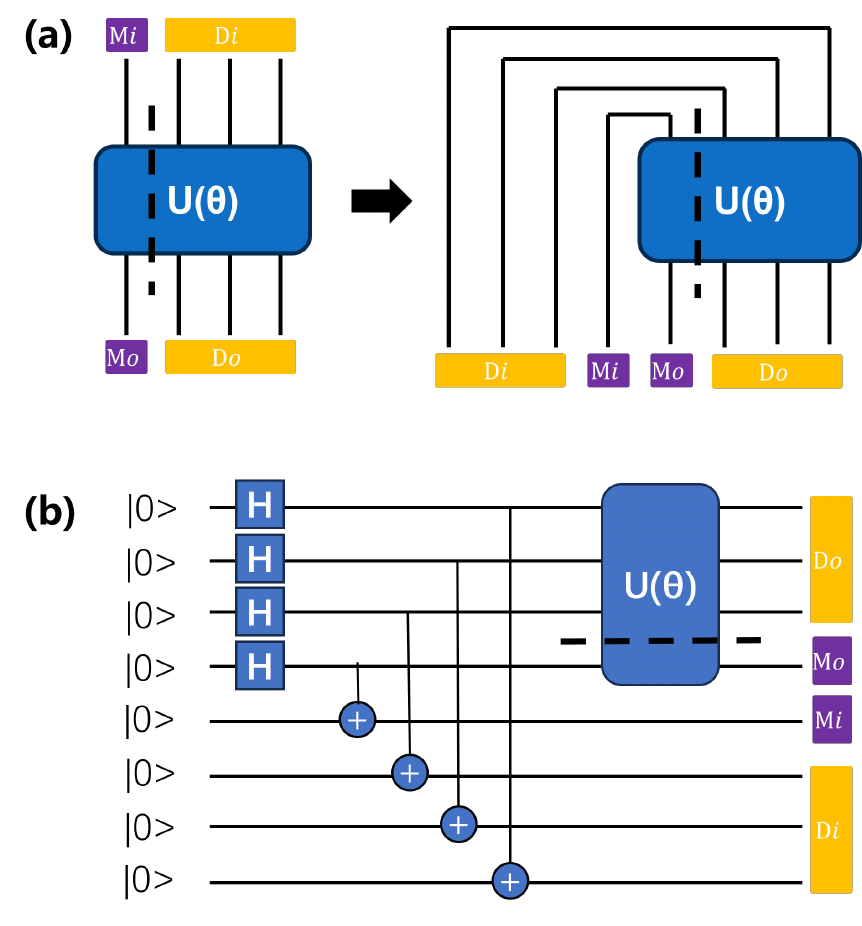}
\caption{(a)Left: We divide the input and output of the $U$ operator into four subsystems depending on whether the qubit has a measurement or not, they are the input of measurement subsystem ($Mi$), the output of measurement subsystem ($Mo$), the input of discard subsystem ($Di$), and the output of discard subsystem ($Do$). Right: In order to facilitate calculation, flipping the input leg of the $U$ operator to down. (b)The schematic diagram of flipping the input leg by EPR pair. }
\label{fig:fig3}
\end{figure}

This paper is inspired by the Scrambling model \cite{15}, the $U$ operator (QNNs Ansatz) is divided into two subsystems, which are measurement subsystem and discarder subsystem, the measurement subsystem is composed of qubits that perform measurement operation, the discarder subsystem is composed of other qubits, as shown in Fig. 3(a). By calculating the mutual information between $Mi$ and $Mo$, as well as $Di$ and $Mo$, during training, we expect to find the working mechanism of QNNs.

In the actual calculation, we flip the input leg of the $U$ operator to down, treat them equal, this can be expressed as formula $|\Psi\rangle=\sum_{j}\sqrt{p_j}|\psi_j\rangle_{in}\otimes|\phi_j\rangle_{out}$, $\rho=|\Psi\rangle\langle\Psi|$ is the density matrix after evolved, the density matrices of different subsystems can be obtained by the partial trace of $\rho$. The flip can be realized by EPR pair, as shown in Fig. 3(b). Suppose the density matrix is $\rho$ after $U$ evolved (we can get it by quantum states tomograph on physical quantum computer), thus the density matrix of $Di$ can be represented as $\rho_{Di}=Tr_{Do,Mo,Mi}(\rho)$, the entropy can be calculated by $S(Di)=-tr(\rho_{Di} log(\rho_{Di}))$, the entropy of subsystem of $Mi$, $Mo$, $DiMo$ and $MiMo$ can be calculated by same method. Thus the mutual information intra- and inter- subsystems can be calculated by the following formula.

\begin{eqnarray}\label{ME1}
&&I(Di:Mo)=S(Di)+S(Mo)–S(DiMo)
\end{eqnarray}
\begin{eqnarray}\label{ME1}
&&I(Mi:Mo)=S(Mi)+S(Mo)–S(MiMo)
\end{eqnarray}

\section{Result and analysis}

\begin{figure*}
\includegraphics[width=18cm]{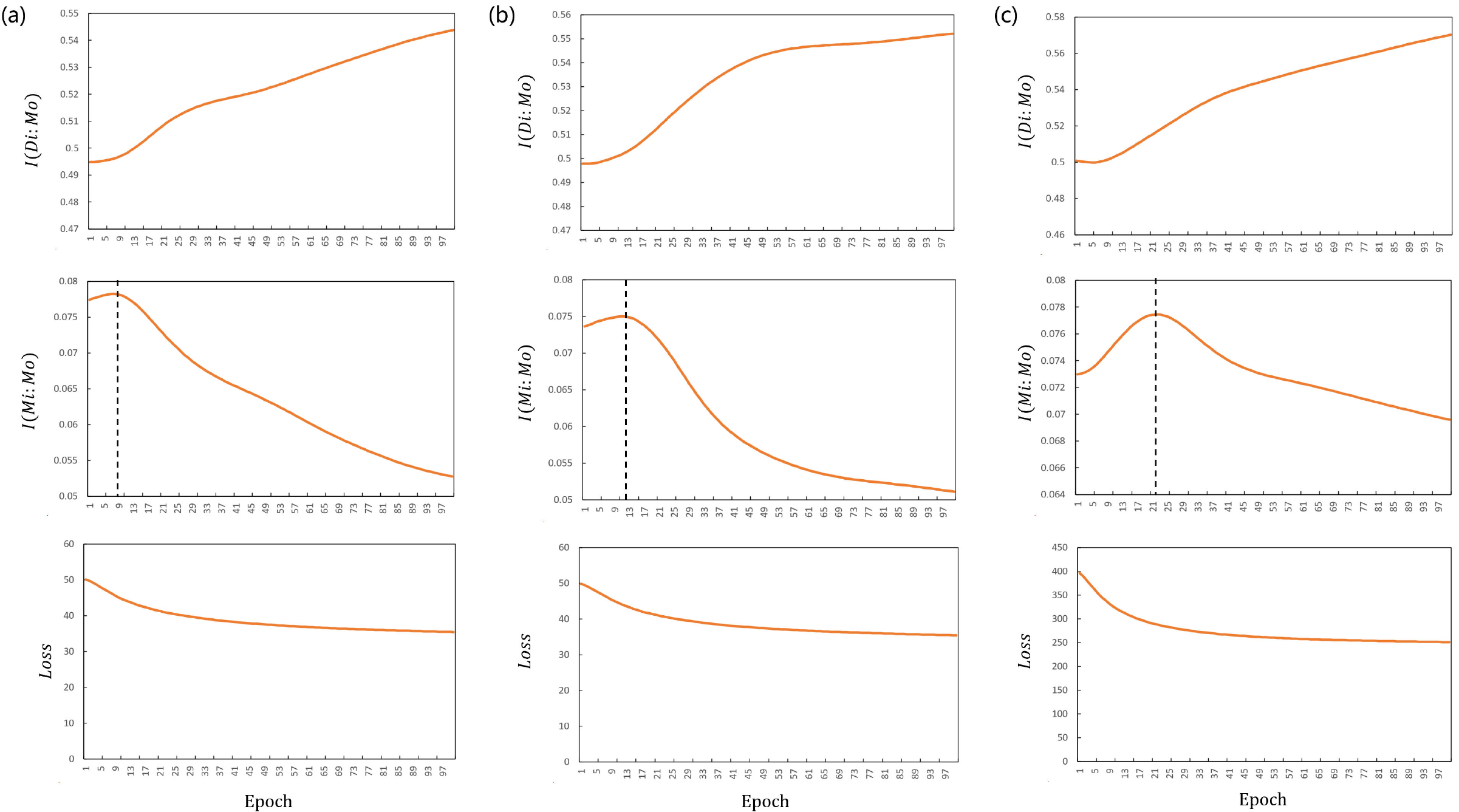}
\caption{The results are averaged over 500 experiments. (a)Result for Iris dataset, Top: $I(Di:Mo)$ as functions of the training epoch, it increase with iteration. Middle: $I(Mi:Mo)$ as functions of the training epoch, which demonstrates a two-phase behavior, it increases before 8 iteration, then begins to decrease. Bottom: Loss as functions of the training epoch, it decrease during training. (b)Result for diabetes dataset, Top: $I(Di:Mo)$ as functions of the training epoch, it increase with iteration. Middle: $I(Mi:Mo)$ as functions of the training epoch, which demonstrates a two-phase behavior, it increases before 12 iteration, then begins to decrease. Bottom: Loss as functions of the training epoch, it decrease during training. (c)Result for Breast Cancer Wisconsin (Original) dataset, Top: $I(Di:Mo)$ as functions of the training epoch, it increase with iteration. Middle: $I(Mi:Mo)$ as functions of the training epoch, which demonstrates a two-phase behavior, it increases before 22 iteration, then begins to decrease. Bottom: Loss as functions of the training epoch, it decrease during training. }
\label{fig:fig4}
\end{figure*}

In this paper, we experimented the explicit quantum models with Brick Wall Ansatz, with $n$=4 (qubit number) and $l$=4 (repeat number). We conducted experiments on four datasets: Iris \cite{29}, diabetes, and Breast Cancer Wisconsin (Original). For Iris dataset, we only selected two categorical of them (setosa and versicolor), and constructed a two-categorized dataset, we use qubit encoding for this dataset. For diabetes dataset, it have 8 features, we use "interleaved" qubit encoding. For Breast Cancer Wisconsin (Original) dataset, we use amplitude encoding, this dataset have 9 features, but the quantum states need encode 16 amplitude, the rest is denoted by 0. The experiments were performed on LFQAP platform \cite{30}, we update the parameters use Adam optimizer, and the total training was performed for 100 iterations. We calculated $I(Di:Mo)$, $I(Mi:Mo)$ and Loss separately for the three datasets, the results are shown in Fig. 4.

Let's look at $I(Di:Mo)$ first, the result is shown in the top of Fig. 4, it is obvious that $I(Mi:Mo)$ increasing with training iteration for the three datasets, this means some information of discard subsystem pushed into the measurement subsystem. In conjunction with the bottom of Fig. 4, the loss decreases during training, so we have strong reasons to assume that the information about label of discard subsystem pushed into the measurement subsystem during training, this caused the loss decreasing.

Next, we focus on $I(Mi:Mo)$, which is shown in the middle of Fig. 4, all three datasets with different encoding demonstrate a two-phase behavior, this is consistent with the information bottleneck theory\cite{12}, so we can assume that the measurement subsystem performs feature fitting in the first phase, and the discard some information in the second phase, because of the loss is decreasing in the whole phase, so the discarded information is not related to label, and it's obvious that the discarded information is pushed into the discard subsystem because of the information conservation.

\begin{figure}
\includegraphics[width=7.5cm]{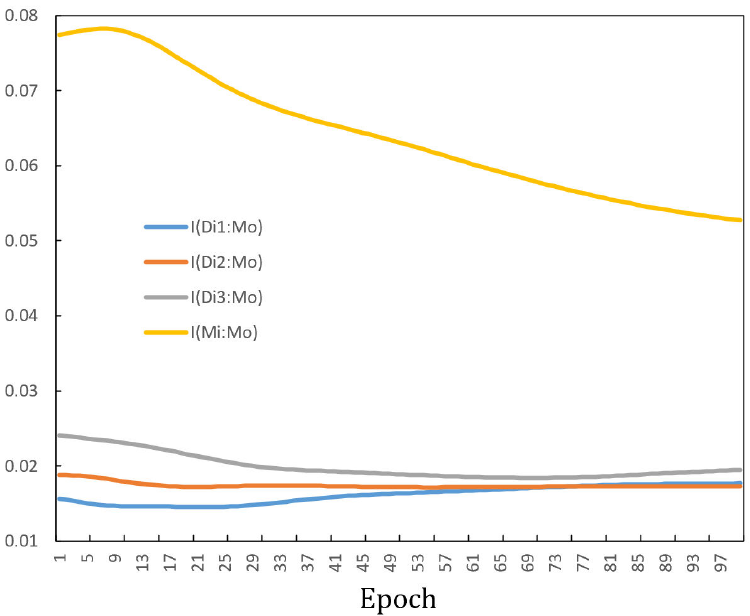}
\caption{The mutual information of $Mo$ and single input qubits of Iris dataset. $Di1$, $Di2$, $Di3$ is the qubits from right to left in upper left of Fig. 3. }
\label{fig:fig5}
\end{figure}

In order to further reveal the behavior of QNNs, We calculated the mutual information of $Mo$ and single input qubits ($Di1$, $Di2$, $Di3$, $Mi$) during training, we consider Iris dataset as an example, the result is shown in Fig. 5. The mutual information $I(Di1:Mo)$, $I(Di2:Mo)$, $I(Di3:Mo)$ have completely different behaviors with $I(Di:Mo)$, and  $I(Di1:Mo) + I(Di2:Mo) + I(Di3:Mo)$ is much smaller than $I(Di:Mo)$, this suggests that quantum entanglement carries the main classification discriminant information in QNNs learning, this may implies that the global features with entanglement more likely to be real discriminative features with generalization ability, the localized feature more likely to be spurious features only valid on specific samples. In the other aspect, $I(Di1:Mo)$, $I(Di2:Mo)$ and $I(Di3:Mo)$ demonstrates a broadly similar behavior, and it is very different with $I(Mi:Mo)$, this further explains the measurement subsystem and discard subsystem have different working mechanisms, it confirms the rationality of the system division.

This experiment reveals the working mechanism of QNNs from the information perspective, for discard subsystem, the information about label was pushed into the measurement subsystem continued during training. for measurement subsystem, we observed a two-phase behavior, it may the process of feature fitting and information compression.

\section{Discussion and outlook}
This paper studied the mutual information behavior in QNNs, we find the label-related information of discard subsystem is pushed into the measurement subsystem during training, and the find a two-phase behavior of measurement subsystem in training process, which consistent with the information bottleneck anticipation. We are able to understand the working mechanism of QNNs to some extent based on experimental results, it can promote the application of QNNs in some critical areas. Further, it also can be used to analyze the accuracy and generalization based on paper \cite{14}.

But it's worth noting that although some results have been achieved in explaining QNNs by using information theory, there are still many key issues that need to be addressed. For example, calculating the mutual information behavior between different layers of QNNs and drawing the information plane (the paper \cite{12} draws the information plane of a classical neural network). And how to use the results of mutual Information to help design more accurate and more trainable QNNs.

\section*{Acknowledgments}
The authors thank Pengfei Zhang for helpful discussions of experimentation.


\begin{thebibliography}{26}
\bibitem{30}
LFQAP is our self-developed quantum artificial intelligence experimental platform.
\end{thebibliography}
\end{document}